\documentclass[aps,prc,twocolumn,showpacs,amsmath,amssymb,nofootinbib]{revtex4}
\usepackage {lscape}
\usepackage{graphicx}
\usepackage{amsmath}
\usepackage{color}

\usepackage{epsfig,colordvi}

\usepackage{graphicx}
\usepackage{dcolumn}
\usepackage{bm}
\usepackage{epsfig}
\usepackage{array}
\usepackage{amsmath}
\usepackage[bookmarks=true]{hyperref}
\usepackage{color}

\hypersetup{
	colorlinks=true,
	pdfborder={0 0 0},
	citecolor=blue,
	linkcolor=blue,
	urlcolor=blue
}

\begin{document}

\title{ Shell model results for $^{47-58}$Ca isotopes in the $fp$, $fpg_{9/2}$ and $fpg_{9/2}d_{5/2}$ model
spaces}

\author{ Bharti Bhoy$^{1}$\footnote{E-mail address: charubharti14@gmail.com}, 
Praveen C. Srivastava$^{1}$\footnote{Corresponding author: praveen.srivastava@ph.iitr.ac.in} and
Kazunari Kaneko$^{2}$\footnote{E-mail address: kkaneko5319@gmail.com}}
\address{$^{1}$Department of Physics, Indian Institute of Technology Roorkee, Roorkee
247 667, India}
\address{$^{2}$Department of Physics, Kyushu Sangyo University, Fukuoka 813-8503, Japan}

\date{\hfill \today}

\begin{abstract}
We have reported shell-model results for $^{47-58}$Ca isotopes in the $fp$, $fpg_{9/2}$ and $fpg_{9/2}d_{5/2}$ model
spaces using realistic $NN$ interaction. We have also performed a systematic shell-model study using interactions
derived from in-medium similarity-renormalization group (IM-SRG) targeted for a particular nucleus with chiral $NN$ and $3N$ forces. The results obtained are in a reasonable agreement with the available experimental data in $fp$ model space with $NN$ interaction.
It is shown that the $g_{9/2}$ and $d_{5/2}$ orbitals play an important role for heavier neutron-rich $^{54-58}$Ca isotopes, while it is marginal for $^{47-52}$Ca.  
We have also examined spectroscopic factor strengths using $NN$ and $NN+3N$ interactions for recently available experimental data.

\end{abstract}

\maketitle

\section{INTRODUCTION}
The study of neutron-rich calcium isotopes is a topic of ongoing  interest to understand the shell evolution and the location
of drip line \cite{nature_ruiz,holtdrip,nadya}.
The discovery of $^{60}$Ca and implication for the stability of $^{70}$Ca has  been recently reported by RIKEN experimental group in the Ref. \cite{022501}.  
In contrast to $ab~initio$ calculations including three-body forces and continuum effects predict that $^{59}$Ca \cite{032502,forssen} is unbound and $^{60}$Ca marginally 
bound and unbound \cite{132501}. The mass measurement of $^{55-57}$Ca \cite{022506} confirmed the $N$=34 subshell closure in $^{54}$Ca. In the recent experiment, 
the robust characteristic of $N=34$ subshell closure has been reported in $^{52}$Ar \cite{072502}.

The neutron-rich Ca isotopes have been previously investigated by the shell model with $NN$ and $NN+3N$ interactions in $fp$ and $fpg_{9/2}$ model spaces \cite{Holt2}.
  Shell-model calculations show that $fpg_{9/2}$ model space can reproduce reasonable spectra 
 up to N $\leq$ 35 but fails to explain strong collectivity in nuclei around $N=40$.
 To reproduce the enhanced  collectivity,  $d_{5/2}$ orbital should be included in $fpg_{9/2}$ model space, because collective behavior can be understood in terms of quasi-SU(3) \cite{Lenzi}. The importance of the 
 $d_{5/2}$ orbital is also reported in Ref. \cite{keneko}.
It has been proposed \cite{Caurier} that for neutron-rich $fp$-shell nuclei, the neutrons are excited to the $sdg$ orbitals coupled to
the unfilled $f_{7/2}$ proton orbital is responsible  for a new region of deformation. Recently, the shell-model interpretation of the first spectroscopy
of $^{61}$Ti using  LNPS interaction for $fpg_{9/2}d_{5/2}$ model space has been reported by Wimmer {\it et al.} in Ref. \cite{wimmer}.
It has been shown that the ground state configuration is dominated by particle-hole excitations to the $g_{9/2}$ and $d_{5/2}$ orbitals.
Thus, in the neutron-rich $fp$ shell nuclei, the inclusion of $g_{9/2}$ and $d_{5/2}$ orbitals in the model space becomes
crucial as we approach
towards $N=40$. 

Earlier, it has been shown that the many-body perturbation theory (MBPT) with three-nucleon forces ($3N$) is very important to explain
 the spectroscopy of neutron-rich  Ca isotopes \cite{Holt1}. In addition, the $ab~initio$ calculations with other modern
approaches: in-medium similarity renormalization group (IM-SRG)  and coupled-cluster effective interaction (CCEI) with
chiral $NN$ and $3N$ forces among valence nucleons are found to describe well the location of drip line \cite{Stroberg}.

The neutron-rich calcium isotopes  are particular attraction for investigating the shell formation. 
 The importance of $3N$ forces are
crucial for explaining spectroscopy of Ca chain as reported in Ref.  \cite{Holt2}.

 Motivated with recent experimental data for spectroscopic
factor strengths for Ca isotopes, we perform shell-model calculations with $NN$ and $NN+3N$ interactions.
The aim of the present manuscript is to investigate recently available experimental data for 
spectroscopy and nuclear observables for the Ca isotopes using shell-model calculations with $NN$ interaction for 
$fp$, $fpg_{9/2}$, and $fpg_{9/2}d_{5/2}$ model spaces.  We have also reported shell model results with $NN+3N$ interaction 
for $fp$ space.
The present study will add more information to earlier theoretical work reported in Refs. \cite{Holt1,Holt2}. 

This paper is organized as follows. In Sec. II, we present details of theoretical formalism. Comprehensive discussions are reported in  Sec. III.
Finally, a summary and conclusions are drawn in Sec. IV.

 \section{THEORETICAL FRAMEWORK}
 
 We can express the present shell-model effective Hamiltonian in terms of single-particle energies and two-body matrix elements numerically,
 \begin{eqnarray}
 	\nonumber H&=&\sum_{\alpha}\varepsilon_{\alpha}{\hat N}_{\alpha} \\
 	&&+ \frac{1}{4}\sum_{\alpha\beta\delta\gamma JT}\langle j_{\alpha}j_{\beta}|V|j_{\gamma}j_{\delta}\rangle_{JT}A^{\dag}_{JT;j_{\alpha}j_{\beta}}A_{JT;j_{\delta}j_{\gamma}},
 	\label{eqa:1}
 \end{eqnarray}
 
 where $\alpha=\{nljt\}$ denote the single-particle orbitals and $\varepsilon_{\alpha}$ stand for the corresponding single-particle energies. 
 $\hat{N}_{\alpha}=\sum_{j_z,t_z}a_{\alpha,j_z,t_z}^{\dag}a_{\alpha,j_z,t_z}$ is the particle number operator. $\langle j_{\alpha}j_{\beta}|V|j_{\gamma}j_{\delta}\rangle_{JT}$ are the two-body matrix elements coupled to spin $J$ and isospin $T$. $A_{JT}$ ($A_{JT}^{\dag}$) is the fermion pair annihilation (creation) operator.

In the present work, we perform shell-model calculations in ${fp}$, ${fpg_{9/2}}$, and ${fpg_{9/2}d_{5/2}}$ model spaces without any truncation.
To diagonalize the matrices, the shell model code KSHELL \cite{Kshell} has been used. The maximum dimension we have diagonalized in the case of $^{58}$Ca for the ground state is 2.7 x 10$^{8}$.

We have taken GXPF1Br+V$_{MU}$ interaction \cite{Tomoaki} for all the three model spaces. Since the GXPF1Br+V$_{MU}$ interaction is made for $fpg_{9/2}d_{5/2}$ model space,
thus while doing calculation for $fp$ and ${fpg_{9/2}}$ model spaces, we allow valence neutrons to occupy in the $f_{7/2}$, $p_{3/2}$, $f_{5/2}$, $p_{1/2}$ orbitals, and
 further including $g_{9/2}$ orbital, respectively. 
  To see the impact of modified single-particle energies on the higher mass side of Ca isotopes,
 we have also reported shell model results for $fpg_{9/2}$ model space with modified single-particle energy of $g_{9/2}$ orbital 
by increasing it with 2 MeV, corresponding results are shown in figures at the last column. 
This part of the calculation is denoted  by $({fpg_{9/2}})_n$.
The $fp$-shell matrix elements are taken from GXPF1Br \cite{steppenbeck}. 
The GXPF1Br \cite{steppenbeck} interaction is modified version of GXPF1B \cite{Honma2} with correction in monopole interaction
for $<0f_{5/2}1p_{3/2}|V|0f_{5/2}1p_{3/2}>_{T=1}$.
The  GXPF1B interaction \cite{Honma2} is upgraded version with the modification of five $T$ = 1 two-body matrix elements and
the bare single-particle energy which involve the 1$p_{1/2}$ orbital from GXPF1A \cite{Honma1}.
The cross-shell two-body interaction between $fp$ and $gds$-shell orbitals are taken from $V_{MU}$ \cite{vmu}.
In the Hamiltonian ( Eq. \ref{eqa:1}) we have added  $\beta_{c.m.}$$H_{c.m.}$ term as proposed by Gloeckner and
Lawson \cite{GL} to remove the spurious center-of-mass motion due to
the excitation beyond one major shell in the case of $fpg_{9/2}$ and $fpg_{9/2}d_{5/2}$ calculations.
We have taken $\beta_{c.m.}$ = 10. There is no further effect on results of energy levels and occupancy of orbitals by increasing value of $\beta_{c.m.}$.
Further, all the two-body matrix elements are scaled by $(A/42)^{-0.3}$ as the mass dependence. 
We have used the harmonic-oscillator parameter $\hbar \Omega = 41 A^{-1/3}$
for all the calculations. Earlier we have reported the importance of $g_{9/2}$ orbital in the model space for Mn 
isotopes in the Ref. \cite{pcs_mn}.

 Stroberg $\it {et ~ al.}$ \cite{Stroberg} presented a nucleus-dependent valence-space approach using the IM-SRG, which
 is normal ordered with respect to a finite-density reference state $|\Phi\rangle$.  
 This approach adopts a decoupled valence space Hamiltonian in which occupied orbitals are fractionalized.
 The effective Hamiltonian can be expressed in terms of single-particle energies and two and three-body matrix elements, as:  
 \begin{multline}
 H = E_0 +  \sum_{ij}f_{ij}\{a_i^{\dagger}a_j\}
 + \frac{1}{4}\sum_{ijkl}\Gamma_{ijkl}\{a^{\dagger}_ia^{\dagger}_ja_la_k\}\\
 + \frac{1}{36}\sum_{ijklmn}W_{ijklmn} \{a^{\dagger}_ia^{\dagger}_ja^{\dagger}_ka_na_ma_l\},
 \label{H}
 \end{multline}
 
 \noindent where $E_0, f_{ij}, \Gamma_{ijkl}$ and $W_{ijklmn}$ are the normal ordered zero-, one-, two-, and three-body terms, 
 respectively.
 The normal ordered strings of creation and annihilation operators obey 
 $\langle\Phi|\{a_i^{\dagger}\ldots a_j\}|\Phi\rangle = 0$. Here chiral $NN$ interaction is taken from N$^{3}$LO \cite{machleidt11_n3lo,machleidt12_n2lo}, and a chiral 3$N$ interaction 
 is taken from N$^{2}$LO \cite{navratil}. 
 To make the calculation easier, the 
 residual 3$N$ interaction $W_{ijklmn}$ is neglected among valence nucleons leading to the normal-ordered two-body approximation.
 


\begin{figure*}
	\vspace{-0.8cm}
	\includegraphics[width=85mm]{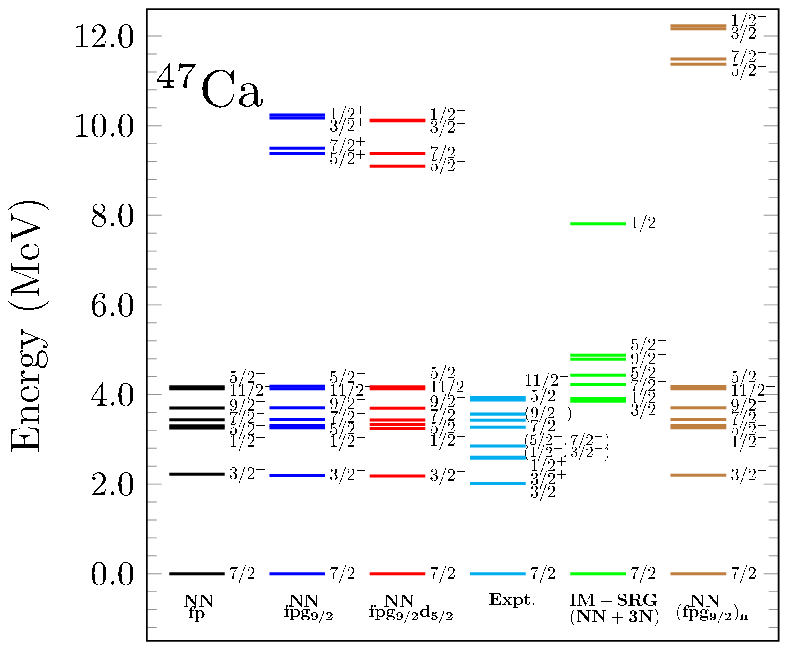}
	\includegraphics[width=85mm]{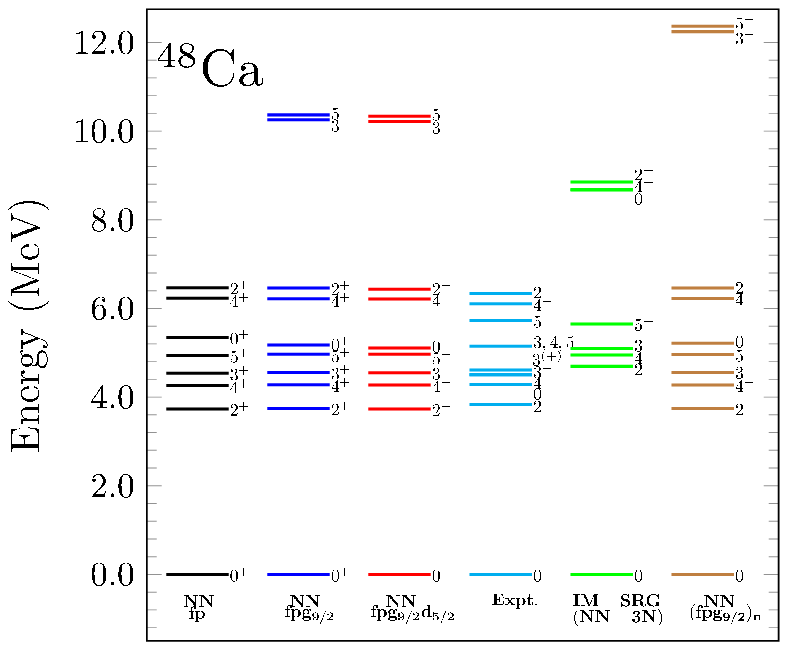}
        \includegraphics[width=85mm]{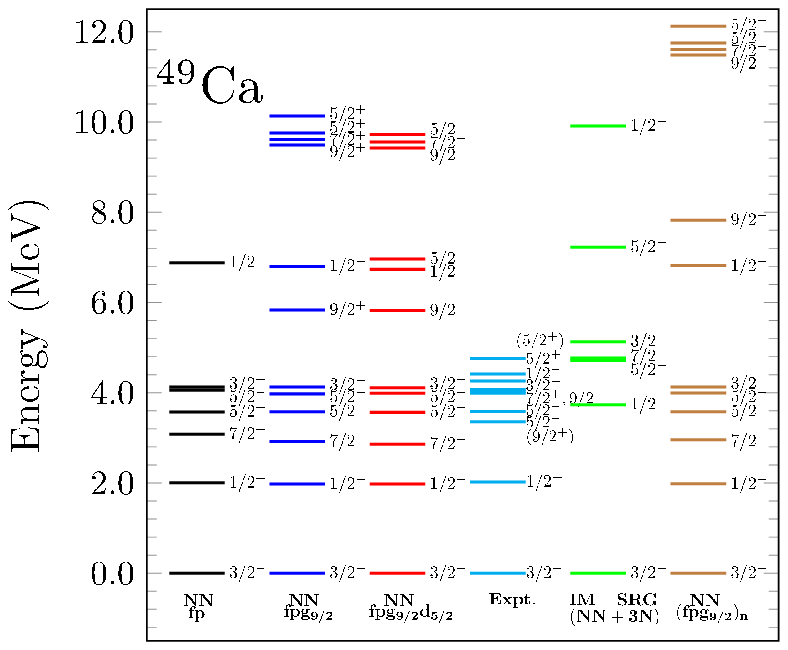}
	\includegraphics[width=85mm]{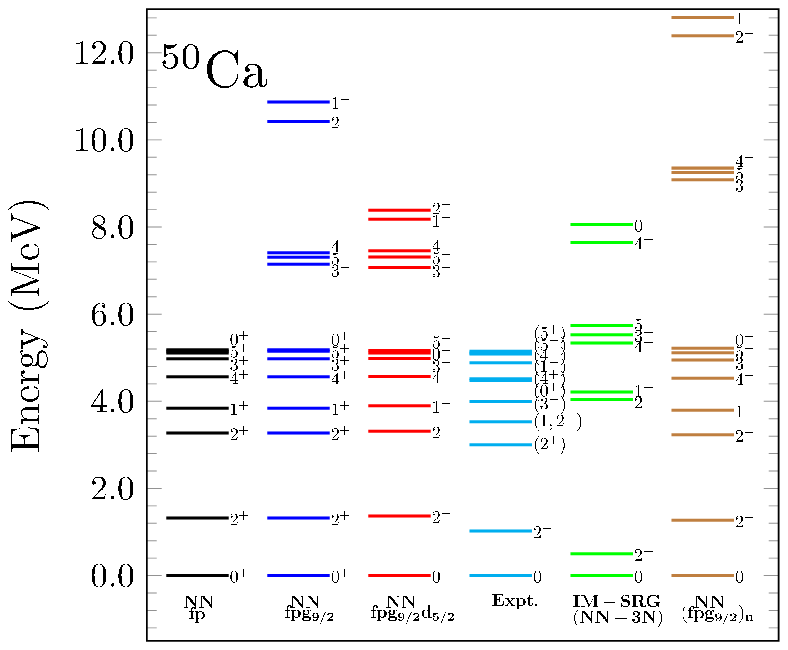}
        \includegraphics[width=85mm]{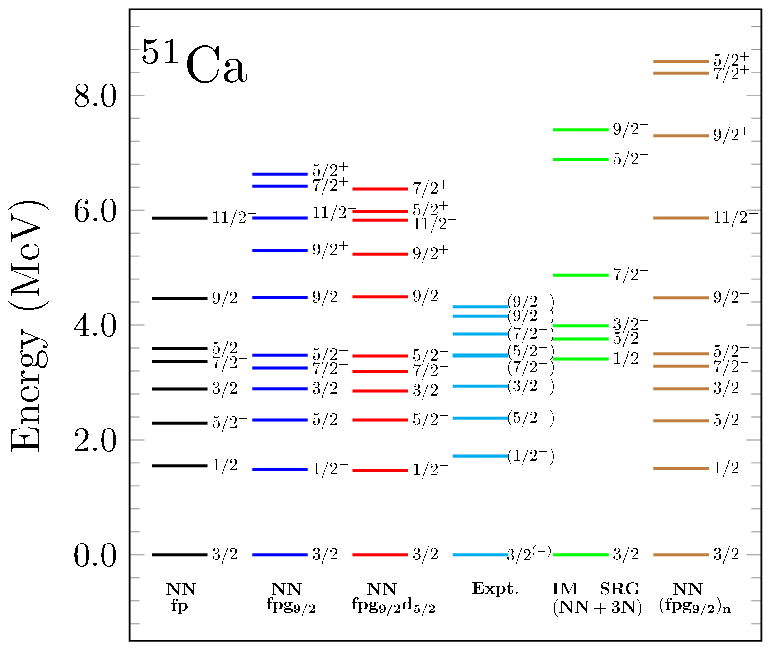}
	\includegraphics[width=85mm]{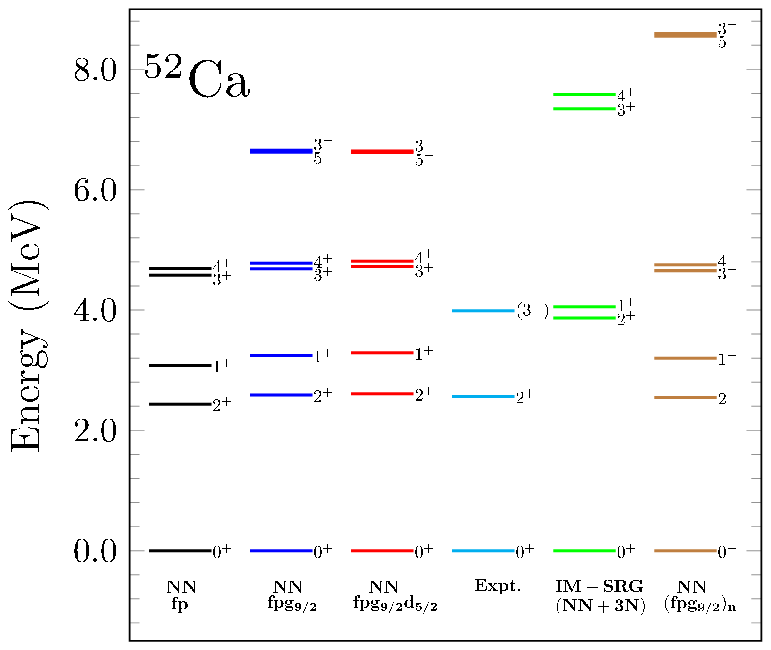}
	\caption{\label{fig1} Comparison between calculated and experimental \cite{NNDC} energy levels for $^{47-52}$Ca.}
\end{figure*}



\begin{figure*}
	\vspace{-1.0cm}
	\includegraphics[width=85mm]{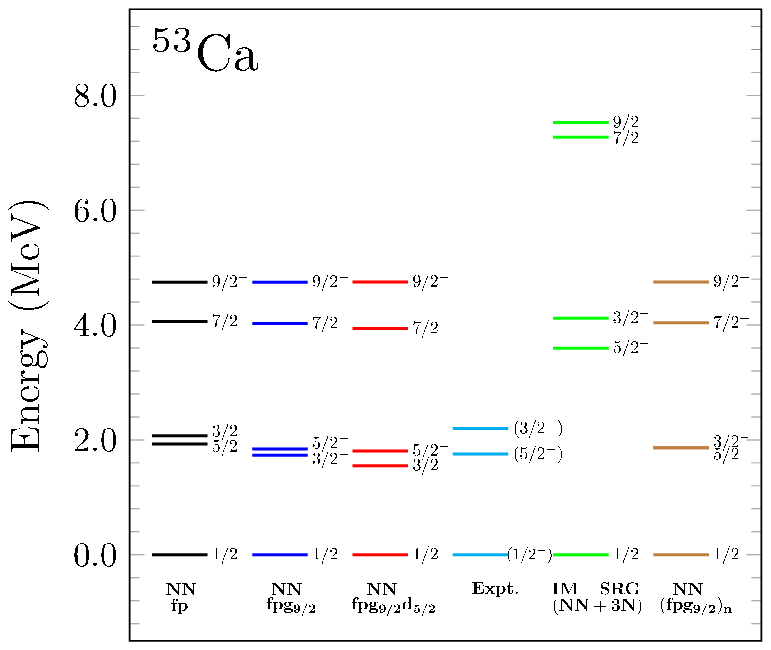}
	\includegraphics[width=85mm]{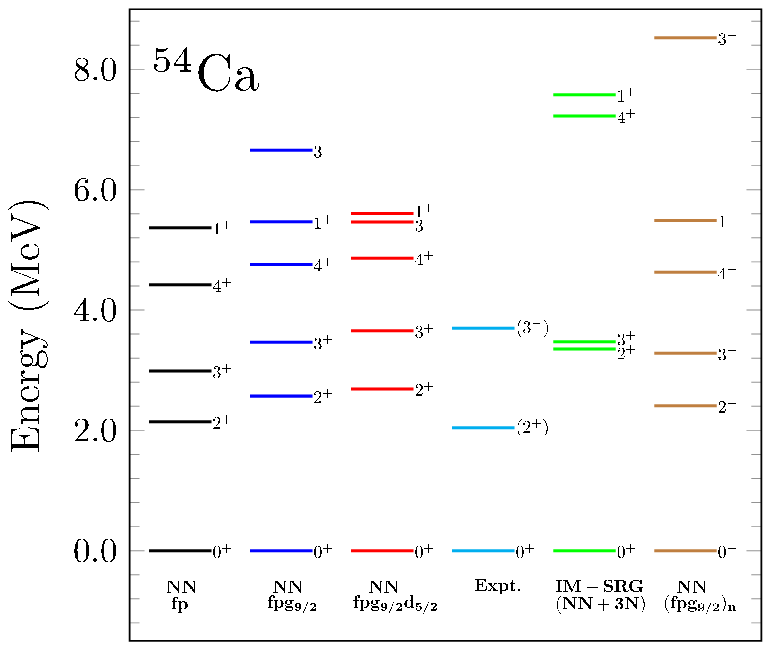}
	\includegraphics[width=85mm]{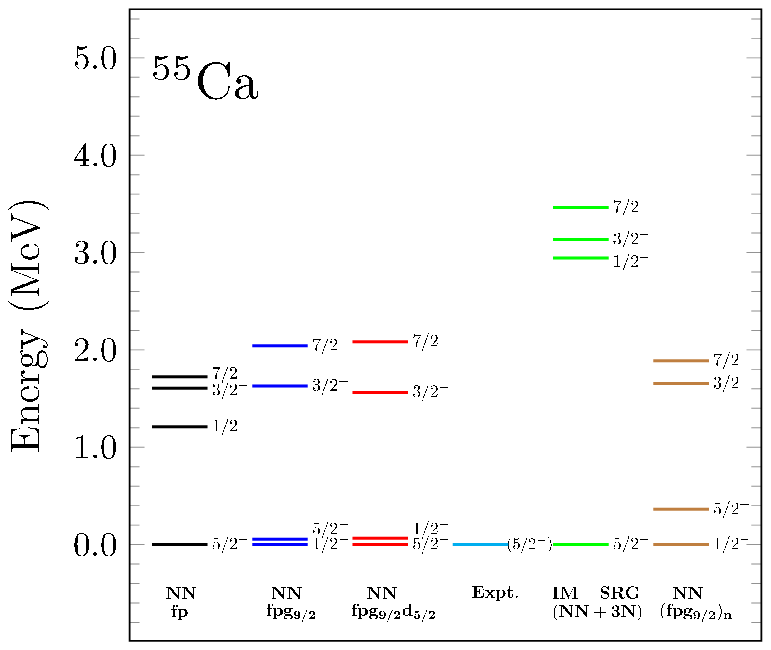}
	\includegraphics[width=85mm]{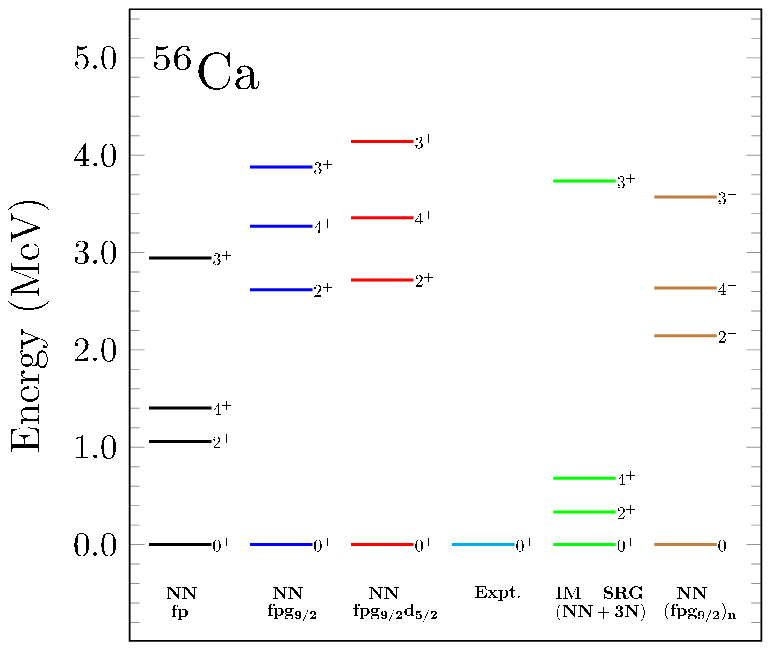}
	\includegraphics[width=85mm]{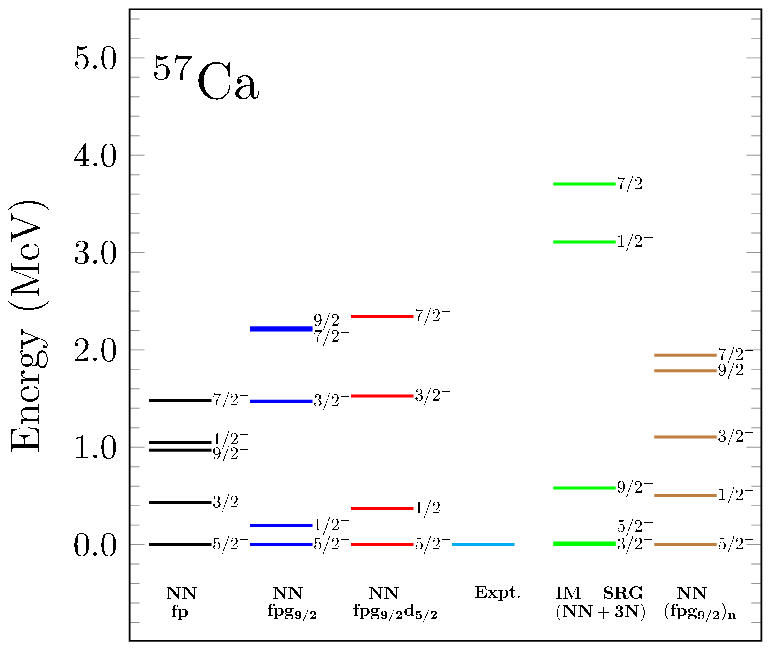}
	\includegraphics[width=85mm]{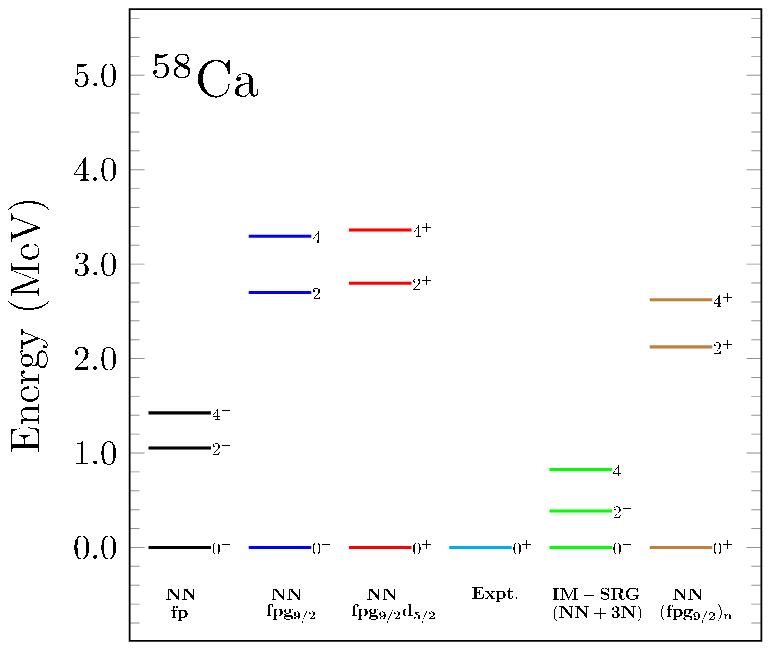}
	
		\caption{\label{fig2} Comparison between calculated and experimental \cite{NNDC} energy levels for $^{53-58}$Ca.}
\end{figure*}

\begin{center}

  \section{RESULTS AND DISCUSSION} 
    \end{center}
The comparisons of energy levels with shell model calculations and experimental data are shown for $^{47-52}$Ca and $^{53-58}$Ca
in Figs. \ref{fig1} and \ref{fig2}, respectively.

  In $^{47}$Ca, the $fp$, ${fpg_{9/2}}$, and ${fpg_{9/2}d_{5/2}}$ model spaces result  for negative parity states are in a reasonable agreement with the experimental data.

  In $^{48}$Ca, the first excited 2$^{+}$ state is higher than those of the neighboring Ca nuclei. All the three set of model space results are in a good agreement 
  with the experimental data for positive parity energy states, while the 3$^{-}$ and 5$^{-}$ states in $fpg_{9/2}$ and $fpg_{9/2}d_{5/2}$ calculation
  are much higher than the experimental data.
 With increasing model space the energy levels are slightly compressing. 

 For $^{49}$Ca, all the calculations with $NN$ force reproduce well first excited $1/2^{-}$ state.
 The ground state in $^{49}$Ca is dominated by the single-particle $p_{3/2}$ state.
 The calculated first $5/2^{+}$ level is at 6.966 MeV in the calculations from $fpg_{9/2}d_{5/2}$ model space and 9.575 MeV with $fpg_{9/2}$ model space.
 For the other excited states, the calculation from all the three  valence spaces reproduce reasonably energy
 levels. In IM-SRG,
 the first excited $1/2^{-}$ state lies at very high energy ($\sim$ 1.7 MeV higher).

 For $^{50}$Ca, the location of the first excited state $2{^+}$ in all the calculations have been predicted very well with experimental data except for IM-SRG result. IM-SRG calculations with $NN + 3N$ forces
 underestimate the first excited state by $~500$ keV. Most of the experimental
 levels are tentative in case of $^{50}$Ca. 
 The large energy difference between the ${2_1}^+$ and ${2_2}^+$ states is reproduced by all $NN$ interactions.
 The spin and parity of the third excited state have not been experimentally identified, but our calculations 
 predict it ${1}^+$ state. 

 In $^{51}$Ca, there is no definite experimental information on the spins and parities of the excited states.
 The first excited ${1/2}^-$ state is indicative of the effective $p_{3/2}-p_{1/2}$ gap and is consistent with the experimental tentative spin assignment.

 The experimental evidence of $N$ = 32 subshell closure for $^{52}$Ca was first time reported in Ref \cite{Huck}. 
 The  calculation from all three valence space reproduces well higher $2_{1}^+$ level,
 consistent with the $N=32$ subshell closure. 
 All the calculations predict the second excited state as ${1}^+$, while 
 this state is not yet observed experimentally.
 The observed (3$^-$) state above 2$_{1}^{+}$ is tentative.
 
 In $^{53}$Ca, the ground state is dominated by $p_{1/2}$ hole. Hence the
 difference between the first excited ${5/2}^-$ and  ${3/2}^-$ levels
 will be mostly due to effective $p_{1/2}-f_{5/2}$ and $p_{1/2}-p_{3/2}$ gaps. 
 This suggests both the $N$ =32 and $N$ = 34 subshell closures. 
 The $fp$,  $(fpg_{9/2})_n$ model space and IM-SRG calculations in $fp$ shell predict ${5/2}^-$ for the first excited state. 
 In the  valence spaces $fpg_{9/2}$ and $fpg_{9/2}d_{5/2}$,
 the calculations predict ${3/2}^-$. 
   The calculation from $(fpg_{9/2})_n$ predicting ${5/2}^-$(1.863 MeV) and  ${3/2}^-$(1.865 MeV) levels as almost degenerate. 
 
 For  $^{54}$Ca, the first experimental spectroscopic study
 on low-lying states was performed  with proton-knockout reactions at RIKEN \cite{steppenbeck}. They observed ${2_1}^+$ state at 2.043 MeV. 
 The calculations predict this state at 2.689 MeV in $fpg_{9/2}d_{5/2}$ space and at 2.569 MeV in $fpg_{9/2}$ space
  and at 2.408 MeV in $(fpg_{9/2})_n$ space. In the IM-SRG calculation,
 the first excited state lies much higher (at 3.352 MeV) than the experimental energy. 
  
  In $^{55-58}$Ca, only spin and parity of ground states are known except for $^{57}$Ca. 
  We have calculated a few low-lying states using the shell-model. Our calculated results will be important for upcoming future experiments.
  
In $^{55}$Ca $fp$, $fpg_{9/2}d_{5/2}$, and IM-SRG  predict ${5/2}^-$ as ground state, consistent with experimental data.

 For $^{56,58}$Ca the first excited ${2}^+$ states are at high energy.
 As $^{54}$Ca is a doubly closed shell nucleus, a ${5/2}^-$ state is expected as a ground state for $^{55}$Ca, with excited states at higher energies. 
 This agrees with the results from the $fp$, $fpg_{9/2}d_{5/2}$ and IM-SRG, but contrasts with the $fpg_{9/2}$ model space results.
The $fpg_{9/2}d_{5/2}$ calculation predicts a very high ${2}^+$ state in $^{56}$Ca at about 2.5 MeV,  higher energy than doubly 
magic isotopes $^{52}$Ca and $^{54}$Ca. In $^{58}$Ca, with a ${2}^+$ state around 3 MeV excitation energy.
To overcome this problem we have tried to modify the single-particle energy of $g_{9/2}$ orbital from 0.881 MeV to 2.881 MeV.
We chose this particular energy from a series of different sets of calculation with modifying single-particle energy 
taking reference of ${2}^+$ state in $^{52}$Ca. The calculation from $(fpg_{9/2})_n$ shows that the $g_{9/2}$ orbital is 
crucial for higher mass region of Ca isotopes. With the modification of the single-particle energy value of $g_{9/2}$ 
orbital, the high ${2}^+$ state starts decreasing  from $^{54}$Ca onwards, however, it show negligible effect below $^{54}$Ca.
 The energy levels with IM-SRG is stretched because $NN$ and $3N$ forces for this interaction were not consistently SRG evolved.

\begin{figure}
\begin{center}
 	\vspace{7.0cm}
        \includegraphics[width=75mm]{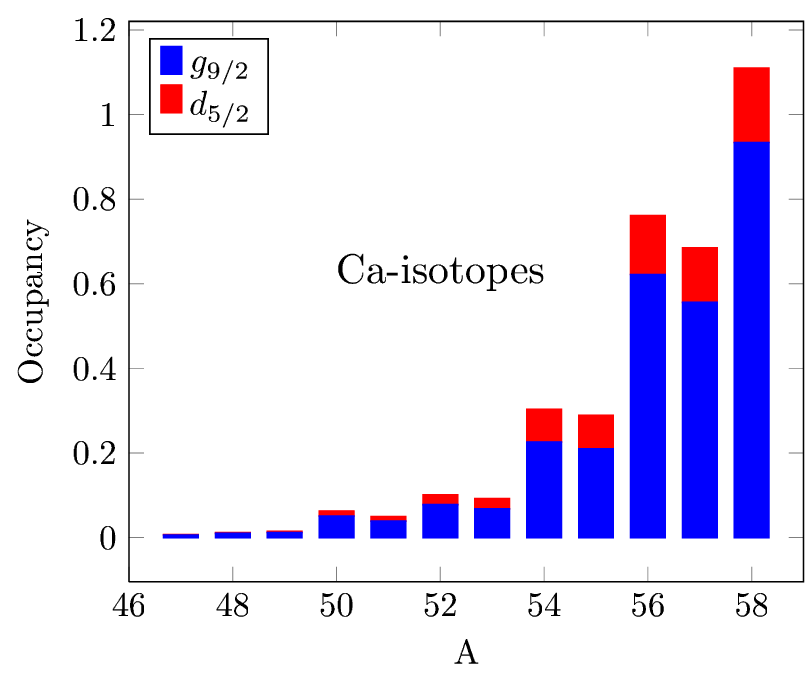}
	\includegraphics[width=75mm]{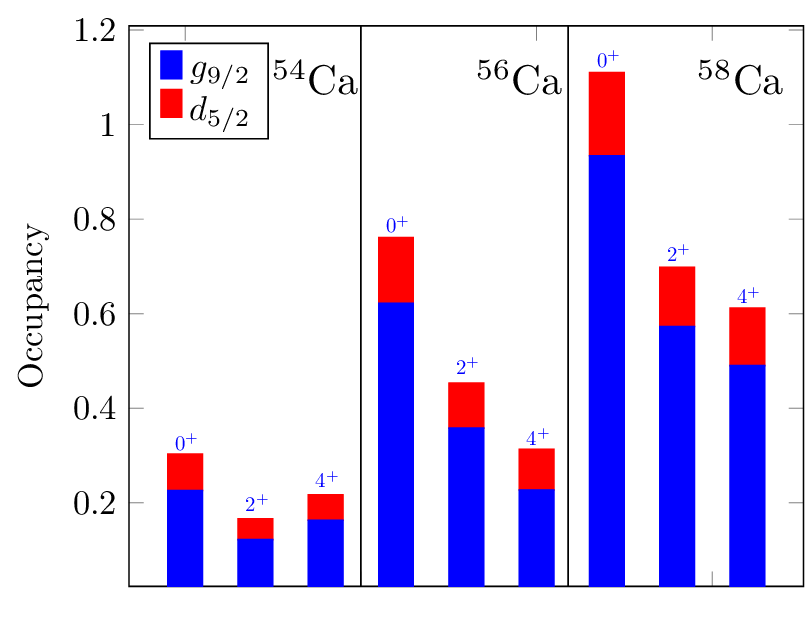}
	\includegraphics[width=75mm]{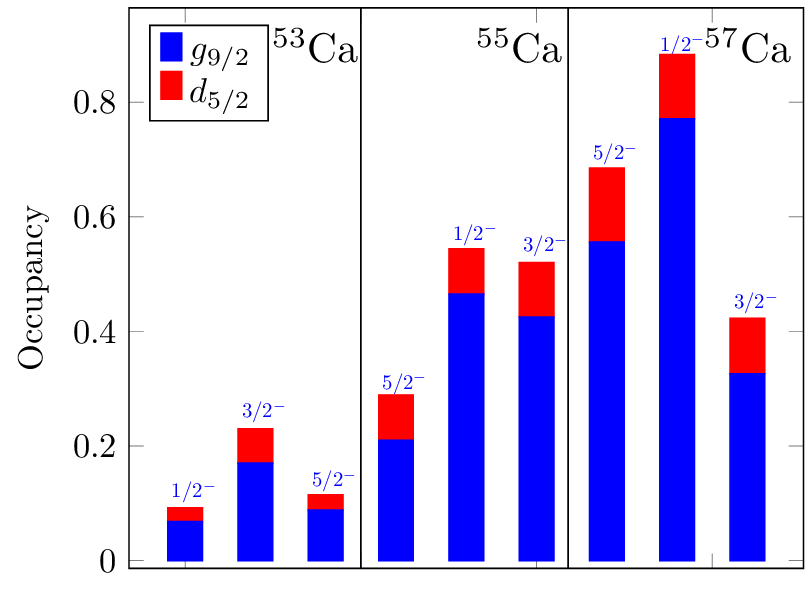}
		\caption{\label{fig_occ} Occupancies of $g_{9/2}$ and $d_{5/2}$ orbitals for the ground and excited states in the Ca isotopes with GXPF1Br+V$_{MU}$ interaction. In the first graph the occupancy of only ground states are shown.}
\end{center}
\end{figure}

  Thus we may conclude that results of $fp$ model space is sufficient to reproduce energy levels of $^{47-52}$Ca isotopes and the role of $d_{5/2}$ orbital
 is very small. Although, with  increasing the neutron number, occupancies of the $g_{9/2}$ and $d_{5/2}$ orbitals increase as shown in the Fig.  \ref{fig_occ}.   The occupancies for $g_{9/2}$
and $d_{5/2}$ decrease for the excited states. This means the effects
of $g_{9/2}$ and $d_{5/2}$  could become important for $B(E2)$ between
the ground state and the first excited state beyond $^{54}$Ca. In fact,
as seen in Table \ref{table1}, the $B(E2)$ increases for $fpg_{9/2}$ and $fpg_{9/2}d_{5/2}$
model space beyond $^{52}$Ca, while they are almost similar to all
model spaces below $^{52}$Ca.

 
 In Table \ref{table1}, we have shown $B(E2)$ values for transitions in Ca isotopes.
 Our calculated results are in a reasonable agreement with the available experimental data. 
 In the calculations, the neutron effective charge is taken as ${e_{n}}$ = 0.5$e$.
 It is clearly seen in Table \ref{table1} that the $g_{9/2}$ and $d_{5/2}$ orbitals affect
 significantly the $B(E2)$ values for heavier $^{54-58}$Ca. Thus the $g_{9/2}$
 and $d_{5/2}$ orbitals play an important role for $^{54-58}$Ca.

\begin{table}
	\caption{\label{table1} $B(E2)$ value in calcium isotopes compared with experiment \cite{montanari,Raman}. 
	The $B(E2)$ values are calculated with IM-SRG and GXPF1Br+V$_{MU}$ ($fp$, $fpg_{9/2}$, and $fpg_{9/2}d_{5/2}$ model spaces) interactions. 
	The units are in $e^2$fm$^4$.}
	\begin{tabular}{cccccccc}
		\hline \hline
		Nuclei & Transition   &\hspace{-1.0cm} Expt.  & \hspace{-1.0cm}\text{$fp$}& \hspace{-0.5cm}\text{$fpg_{9/2}$} &\hspace{-0.40cm} \text{$fpg_{9/2}d_{5/2}$} & IM-SRG \\ 
		\hline
		& & &  &    \\
		
		$^{47}$Ca & $3/2^{-}\rightarrow7/2^{-}$ & 4.0 $\pm$ 0.2  & 3.22 & 3.08 & 3.02  & 1.58  \\			
		$^{48}$Ca & $2^{+}\rightarrow 0^{+}$ & 19 $\pm$ 6.4  & 10.35 & 10.50 & 10.60 & 11.82 \\
		$^{49}$Ca & $7/2^{-}\rightarrow $3/2$^{-}$&   0.53 $\pm$ 0.21  &3.53 & 3.27 & 3.28 &  0.001 \\
		
		$^{50}$Ca & $2^{+} \rightarrow 0^{+}$ & 7.4 $\pm$ 0.2  & 7.82 & 7.82 & 8.01 & 8.0  \\
		$^{51}$Ca & $7/2^{-}\rightarrow $3/2$^{-}$ & N/A  & 6.72 & 6.44 & 6.43 &  7.64 \\
		$^{52}$Ca & $2^{+} \rightarrow 0^{+}$ & N/A  & 6.16 & 6.67 & 7.06  & 6.46 \\
		$^{53}$Ca & $7/2^{-}\rightarrow $3/2$^{-}$ & N/A  & 4.48 & 5.28 & 6.10 & 2.09  \\
		$^{54}$Ca & $2^{+} \rightarrow 0^{+}$ & N/A  & 6.34 & 7.95 & 8.55 &  6.13 \\
		$^{55}$Ca & $7/2^{-}\rightarrow $3/2$^{-}$ & N/A  & 5.39 & 2.69 & 2.02 & 4.55  \\
		$^{56}$Ca & $2^{+} \rightarrow 0^{+}$ & N/A  & 8.95 & 13.15 & 13.92 & 6.90  \\
		$^{58}$Ca & $2^{+} \rightarrow 0^{+}$ & N/A  & 8.25 & 10.48 & 11.12 & 6.85  \\

		\hline \hline
	\end{tabular}
\end{table}

 In Table \ref{table2}, we present the calculated spectroscopic quadrupole moments and magnetic moments
for odd-mass of calcium isotopes using  
GXPF1Br+VMU ($fp, fpg_{9/2}, fpg_{9/2}d_{5/2}$ model spaces) and IM-SRG ($fp$) interactions.
The overall calculated results are in good agreement with the experimental data for magnetic moments.
For $^{53,55,57}$Ca, the experimental data are not available.
For $^{47,49,51}$Ca isotopes, the single particle magnetic moment value is -1.913 corresponding to the last filled neutron in $f_{7/2}$ ($^{47}$Ca) and $p_{3/2}$ ($^{49,51}$Ca) orbitals. The calculated magnetic
moments of $^{47}$Ca and $^{49}$Ca are somewhat close to the effective single particle moment, indicating less contribution of the higher orbitals. For $^{51}$Ca the difference in the single particle magnetic moment and theoretical calculation is large, showing collective effect of orbitals. For the $^{53,55,57}$Ca isotopes, the single particle magnetic moment corresponding to the last occupied orbital $f_{5/2}$ is +1.366. For $^{53}$Ca, the magnetic moment value is very less than the single particle magnetic moment, this shows that the ground state for this isotope has a mixed configuration. 
In the case of $^{55}$Ca for $fpg_{9/2}$ interaction there is a drop in magnetic moment value, which maybe related to the ground state prediction, here we have shown value for the ${5/2}^-$ state, for the obtained ground state ${1/2}^-$ (from SM) also magnetic moment value is very small 0.2249. The magnetic moment values obtained from all the interactions are almost same for $^{47,49,51}$Ca isotopes. For $^{57}$Ca  calculated magnetic moment value becomes close to single particle magnetic moment for the $fpg_{9/2}d_{5/2}$ interaction.
 In the case of $^{57}$Ca the calculated magnetic moment for 5/2$^-$ state is larger with $fpg{_9/2}d_{5/2}$ model space in comparison to $fp$ and $fpg_{9/2}$ model spaces. This reflects the effect of inclusion of $d_{5/2}$ orbital in the model space.

The calculated spectroscopic quadrupole moments are also compared to the known
experimental values. Here we can see, a good description of the
data has been obtained from all the interactions. The value of single particle quadrupole moment (in $eb$) for $^{47,49,51,53,55,57}$Ca isotopes are 0.075, -0.046, 0.047, -0.049, -0.071, 0.073, respectively. Since single particle quadrupole moments are not very close to the experimental quadrupole moment for $^{47,49,51}$Ca,  thus we may conclude that the single particle contribution is not very strong for these isotopes, although there is a small configuration mixing. The IM-SRG interaction values are much closer to the single particle quadrupole moment. For $^{55}$Ca and $^{57}$Ca isotopes calculated quadrupole moments are very less than the single particle quadrupole moments.

    Next, we study spectroscopic factor strengths $C^{2}S$ associated with neutron-hole states in $^{47-53}$Ca from IM-SRG and GXPF1Br+V$_{MU}$ ($fp$, $fpg_{9/2}$, and 
    $fpg_{9/2}d_{5/2}$ model spaces) interactions.
   Experimental data are available for $^{48}$Ca$\rightarrow$ $^{47}$Ca and $^{50}$Ca$\rightarrow$ $^{49}$Ca transitions. 
   The calculated results are compared with the experimental data \cite{Crawford} in Table \ref{table3}.
   For the $^{48}$Ca$\rightarrow$ $^{47}$Ca transition,  IM-SRG result is $C^{2}S_{th}$ = 7.55 corresponding to observed $C^{2}S_{exp}$ = 6.4$_{-0.9}^{+0.4}$ for the 
   lowest ${7/2}^-$ state, while GXPF1Br+V$_{MU}$ is giving $C^{2}S_{th}$ $\sim$ 7.7. 
   The calculated spectroscopic factor to the first excited ${3/2}^-$ state is too small in comparison with the experimental value. 
   For $^{50}$Ca$\rightarrow$ $^{49}$Ca, 
   the spectroscopic factor from the IM-SRG for the 3/2$^{-}$ and 7/2$^{-}$ states are reasonable as in the experimental data,
   while it is quite small for the 1/2$^{-}$ state. From GXPF1Br+V$_{MU}$ interaction the spectroscopic factor for the 3/2$^{-}$ and 1/2$^{-}$ states
   are reasonable as in the experimental data and giving a large value for 7/2$^{-}$ state.
   For $^{52}$Ca$\rightarrow$ $^{51}$Ca and $^{54}$Ca$\rightarrow$ $^{53}$Ca transitions, we have reported theoretical results of the spectroscopic factor
   for future experiment.
 Recently, the ab initio calculation for $C^{2}S$ corresponding to $^{54}$Ca$\rightarrow$ $^{53}$Ca transition
using N2LO$_{sat}$ and NN + 3N (lnl) interactions are reported in the Ref. \cite{PRL_nav}. In the present work, we have
compared our calculated results with these ab initio results in the Table \ref{table1}. 
The ab initio $C^2S$ \cite{PRL_nav} are lower than the GXPF1 ones because due to collective excitations. Our results for $C^2S$ also  become smaller once we increase model space from $fp$ to  $fpg_{9/2}$ to $fpg_{9/2}d_{5/2}$.

  \begin{table*}
 	\caption{\label{table2}Comparison of experimental \cite{ruiz} and theoretical quadrupole and magnetic moments of ground states. 
 	Shell model results obtained from IM-SRG and GXPF1Br+V$_{MU}$ ($fp$, $fpg_{9/2}$, and $fpg_{9/2}d_{5/2}$ model spaces) interactions.
 	We have taken ${e_{n}}$ = 0.5$e$ and $g_s^{eff}$=$g_s^{free}$. }
 	\begin{tabular}{cccccccccccc}
 		\hline \hline
 		A &  & \multicolumn{2}{c}\text{$\mu(\mu_{N})$ }  &  & & & \multicolumn{2}{c}\text{{Q(eb)}} &\\ \hline
 		
 		\hline	
 		
 		\textbf{} &  $J^\pi$ & Expt.  & \text{$fp$}& \text{$fpg_{9/2}$}& \text{$fpg_{9/2}d_{5/2}$} &  \text{IM-SRG} & Expt.  & \text{$fp$} & \text{$fpg_{9/2}$} & \text{$fpg_{9/2}d_{5/2}$} & \text{IM-SRG} \\
 		\hline
 		
 		47& $7/2^-$ & -1.4064(11) & -1.4618 & -1.4661 & -1.4679 & -1.3640  & +0.084(6)   & +0.0675 & +0.0673 & +0.0675  &+0.0790\\
 		49& $3/2^-$ & -1.3799(8)  & -1.3921 & -1.3919 & -1.3857 & -1.3290  & -0.0360(3)   & -0.0386 & -0.0386 & -0.0386  & -0.0452\\
 		51& $3/2^-$ &-1.0496(11)  & -1.0077 & -1.0503 & -1.0523 & -1.0610 & +0.036(12)  & +0.0390 & +0.0371 & +0.0369  & +0.0421\\
 		53& $1/2^-$ &   N/A    & +0.5063 & +0.5120 &  +0.5020 &  +0.4930 &        & &  &  &   \\
 		55& $5/2^-$ &   N/A    & +1.0687 & +0.3617\footnote{Although $1/2^-$ predicted as a g.s. from SM} &  +1.0316  &+0.9870  &  N/A  &  -0.0550 & -0.0476\footnote{Although $1/2^-$ predicted as a g.s. from SM} & -0.0482 &  -0.0567 \\
 		57& $5/2^-$ &   N/A      & +0.486  & +0.472 &  +1.1668        &   +1.176\footnote{Although $3/2^-$ predicted as a g.s. from SM}  &  N/A  & +0.003     &  +0.001  &   +0.001  &  -0.0012 \footnote{Although $3/2^-$ predicted as a g.s. from SM}\\
 		\hline \hline
 	\end{tabular}
 \end{table*}

\begin{table*}
	\caption{\label{table3} Comparison of experimental \cite{Crawford} and theoretical spectroscopic factor strenghts  obtained from IM-SRG and GXPF1Br+V$_{MU}$ ($fp$, $fpg_{9/2}$, and $fpg_{9/2}d_{5/2}$ model spaces) interactions for different transitions.}
	\begin{tabular}{cccccccccc}
		\hline 
		Level energy (keV)& $J^{\pi}$ & $C^{2}S_{exp}$ & IM-SRG & & {$fp$} & {$fpg_{9/2}$} & {$fpg_{9/2}d_{5/2}$} & N2LO$_{sat}$ \cite{PRL_nav} & NN + 3N (lnl) \cite{PRL_nav}\\ \hline\\
		$^{48}$Ca$\rightarrow$ $^{47}$Ca	 & &  &  &   &  &  \\
		0 & 7/2$^{-}$ & 6.4$_{-0.9}^{+0.8}$ &  7.5 & & 7.7 & 7.6 & 7.6   &  & \\	
		2014 & 3/2$^{-}$ & $\leq$1.4 & 0.002 & & 0.05 & 0.05 & 0.06 &  & \\
		\hline\\
		$^{50}$Ca$\rightarrow$ $^{49}$Ca	 & &  &  &  &  \\
		0 & 3/2$^{-}$ & 2.1(3) & 1.8 & &1.8 &1.7 & 1.6 &  &  \\
		2023 & 1/2$^{-}$ & 0.28$^{+0.05}_{-0.03}$ & 0.04 & & 0.09 & 0.1 & 0.1 &  &  \\
		3357\cite{Crawford} & 7/2$^{-}$ & 3.4$^{+0.4}_{-0.3}$ & 4.7 & &	 7.7 & 7.6 & 7.6 &  &  \\
		\hline\\
		$^{52}$Ca$\rightarrow$ $^{51}$Ca	 & &  & &  &   \\
		0 & 3/2$^{-}$ & N/A & 3.7 & & 3.5 & 3.4 & 3.3  &  & \\
		1718 & 1/2$^{-}$ & N/A & 0.04 & & 0.1 & 0.1 & 0.2 &  &  \\
		2378 & 5/2$^{-}$ & N/A & 0.003 & & 0.003 & 0.01 & 0.01 &  & \\
		\hline\\
		$^{54}$Ca$\rightarrow$ $^{53}$Ca	  & &  & &  &   \\
		0 & 1/2$^{-}$ & N/A & 1.9 & & 1.8 & 1.4 & 1.3  &  1.56 & 1.58  \\
		1738\cite{PRL_nav} & 5/2$^{-}$ & N/A & 0.01 & & 0.1 & 0.5 & 0.5  & 0.01 &  0.02\\
		2220\cite{PRL_nav} & 3/2$^{-}$ & N/A & 3.8 & & 3.5 & 3.4 & 3.4  &  3.12 & 3.17\\

		\hline
	\end{tabular}
\end{table*}

The poor  spectroscopy produced by the IM-SRG  might be due to 
(i) if we look Stroberg et al. \cite{Stroberg}, in the Fig. 2,  the calcium 
isotopes around $^{48}$Ca are overbound by something like 50 MeV. This is due to at 
least in part to the fact that the $NN$ and $3N$ forces for this interaction 
were not consistently SRG evolved. The inconsistent evolution led to the 
oxygen chain agreeing with experiment, but this is almost certainly due to 
a lucky cancellation for that mass region. As we can see, the C isotopes are 
underbound and the Ca and Ni are overbound.
(ii) also inadequacies of the initial
chiral EFT interaction, in the Ref. \cite{gazit} it is reported  that the way that the $3N$ force 
was fit for that interaction was a mistake. Basically, it was fit to the 
beta decay of the triton, but an incorrect conversion between the axial 
current and the $3N$ coupling was used. 
 
 Previously, shell-model calculation for Ca isotopes has been done  in Ref. \cite{Holt2} with many-body perturbation theory (MBPT) using $NN+3N$ forces in the extended model space $pfg_{9/2}$. In our present work, we have tried to get better energy spectra for neutron-rich calcium isotopes along with nuclear observables. As the result reported in Ref. \cite{Holt2}, energy spectra is very much compressed from  $^{47}$Ca to $^{49}$Ca, although after inclusion of $3N$ forces, somehow it improves but still gives approximately 1$\sim$2 MeV compressed spectra. In our calculations up to $^{54}$Ca where experimental data are known, we can obtain very reasonable energy spectra for natural parity states. The calculation from MBPT contrary to experiment giving $0^+$ as the first excited state in $^{48}$Ca and $^{54}$Ca. In the present work, we have $2^+$ as the first excited state in $^{48}$Ca and $^{54}$Ca as in the experimental data at similar energy difference from the g.s. In comparison with the  Ref. \cite{Holt2}, our calculation predicts both $N$=32 and 34 subshell closure very well. Also, the calculated first and second  excited states of $^{51}$Ca and $^{53}$Ca suggest the  subshell closure at $N$=32 and 34. \\

\begin{table*}
	\caption{\label{table4} The calculated shell model wave functions corresponding to $3^{-}$ state in $^{48,50,52,54}$Ca isotopes.}
	\begin{tabular}{ccccc}
		\hline \hline
                Nuclei     & State  & $fpg_{9/2}$  &        $(fpg_{9/2})_n$ & $fpg_{9/2}d_{5/2}$ \\
               \hline \hline \\
		$^{48}$Ca & $3^{-}$ & 91\%  $f_{7/2}^7g_{9/2}^1$ & 91\%  $f_{7/2}^7g_{9/2}^1$ &   90\%  $f_{7/2}^7g_{9/2}^1$\\
  			          &         & 3.4\%  $f_{7/2}^6p_{3/2}^1g_{9/2}^1$& 3.5\% $f_{7/2}^6p_{3/2}^1g_{9/2}^1$ & 3.7\%  $f_{7/2}^6p_{3/2}^1g_{9/2}^1$  \\
              	          &         & 1.5\%  $f_{7/2}^5p_{3/2}^2g_{9/2}^1$ & 1.5\%   $f_{7/2}^5p_{3/2}^2g_{9/2}^1$ &  1.5\%  $f_{7/2}^5p_{3/2}^2g_{9/2}^1$\\
                 \hline \\
		$^{50}$Ca & $3^{-}$ & 93.8\%  $f_{7/2}^8p_{3/2}^1g_{9/2}^1$ & 93.3\% $f_{7/2}^8p_{3/2}^1g_{9/2}^1$ &  87\% $f_{7/2}^8p_{3/2}^1g_{9/2}^1$\\			       
                      &         & 1.4\% $f_{7/2}^6p_{3/2}^3g_{9/2}^1$& 1.4\%  $f_{7/2}^6p_{3/2}^3g_{9/2}^1$ &  2.6\% $f_{7/2}^8p_{3/2}^1d_{5/2}^1$ \\
              	          &         & 1.1\% $f_{7/2}^6p_{3/2}^1f_{5/2}^2g_{9/2}^1$ & 1.1\%  $f_{7/2}^6p_{3/2}^1f_{5/2}^2g_{9/2}^1$ & 1.6\% $f_{7/2}^7p_{3/2}^2g_{9/2}^1$ \\
              \hline \\$^{52}$Ca & $3^{-}$ & 87\%  $f_{7/2}^8p_{3/2}^3g_{9/2}^1$  & 88\% $f_{7/2}^8p_{3/2}^3g_{9/2}^1$ & 83.5\% $f_{7/2}^8p_{3/2}^3g_{9/2}^1$  \\			      
                  &         & 3.3\% $f_{7/2}^8p_{3/2}^1p_{1/2}^2g_{9/2}^1$ & 3.2\%  $f_{7/2}^8p_{3/2}^1p_{1/2}^2g_{9/2}^1$ & 3.7\%  $f_{7/2}^8p_{3/2}^1p_{1/2}^2g_{9/2}^1$ \\
              	          &         & 2.5\% $f_{7/2}^8p_{3/2}^1f_{5/2}^2g_{9/2}^1$ & 2.2\%  $f_{7/2}^8p_{3/2}^1f_{5/2}^2g_{9/2}^1$& 2.6\%  $f_{7/2}^8p_{3/2}^1f_{5/2}^2g_{9/2}^1$  \\
              \hline \\
	$^{54}$Ca & $3^{-}$ & 77.3\%  $f_{7/2}^8p_{3/2}^3p_{1/2}^2g_{9/2}^1$ & 77.5\% $f_{7/2}^8p_{3/2}^3p_{1/2}^2g_{9/2}^1$ & 86.2\% $f_{7/2}^8p_{3/2}^4p_{1/2}^1d_{5/2}^1$ \\			
                   &         & 8.7\% $f_{7/2}^8p_{3/2}^3f_{5/2}^2g_{9/2}^1$ & 9.1\%  $f_{7/2}^8p_{3/2}^3f_{5/2}^2g_{9/2}^1$ &  3.2\% $f_{7/2}^8p_{3/2}^2f_{5/2}^2p_{1/2}^1d_{5/2}^1$  \\
              	          &         & 3.5\% $f_{7/2}^8p_{3/2}^3g_{9/2}^3$ & 2.9\%  $f_{7/2}^8p_{3/2}^3f_{5/2}^1p_{1/2}^1g_{9/2}^1$&  1.2\% $f_{7/2}^6p_{3/2}^4f_{5/2}^2p_{1/2}^1d_{5/2}^1$  \\
              \hline \hline
	\end{tabular}
\end{table*}

\section{Excitation Energies of $3_1^-$ state in even-even Ca isotopes}

The wave function corresponding to $3_1^-$ state in the $^{48,50,52,54}$Ca isotopes 
with $fpg_{9/2}$ and $(fpg_{9/2})_n$ interactions are shown in the Table \ref{table4}.
The calculated $3_1^-$ state
for $^{48,50,52,54}$Ca isotopes with  $(fpg_{9/2})_n$ interaction is higher than that of the $fpg_{9/2}$ interaction due to increase in the single particle energy of $g_{9/2}$ orbital
by 2 MeV.
 The occupancy of  $p_{3/2}$ orbital is changing from 0.073 ($^{48}$Ca) to 1.037 ($^{50}$Ca) to 2.835 ($^{52}$Ca) to 2.943 ($^{54}$Ca) with
 original $fpg_{9/2}$ interaction. 
The $3_1^-$ state is dominated by $\nu (p_{3/2}^1g_{9/2}^1)$ configuration.

Since the $(fpg_{9/2})_n$  interaction predicts $3_1^-$ state at 12.244 MeV for $^{48}$Ca, the calculated $J^\pi$ = $5/2^+$, $7/2^+$ and $9/2^+$ states around 12 MeV in $^{49}$Ca are the multiplets coming from coupling of ($3_1^-$$\otimes$$p_{3/2}^1$)$_J$. Similarly  $(fpg_{9/2})_n$  interaction predicts $3_1^-$ state at 9.085 MeV for $^{50}$Ca, the multiplets of ($3_1^-$$\otimes$$p_{3/2}^1$)$_J$ is responsible for $J^\pi$ = $5/2^+$, $7/2^+$ and $9/2^+$ states in $^{51}$Ca.

\section{CONCLUSIONS} 

In the present work, we have performed shell-model calculations with realistic $NN$
interactions for $^{47-58}$Ca isotopes.
 To see the importance of $g_{9/2}$ and $d_{5/2}$ orbitals, we have performed calculations
for $fp$, $fpg_{9/2}$, and $fpg_{9/2}d_{5/2}$ model spaces. 
In our calculations, after $^{54}$Ca, we are getting a very high ${2}^+$ state in the case of $^{56,58}$Ca.
Thus, to reduce the energy of  ${2}^+$ state and to see the importance of $g_{9/2}$ orbital,
we have increased the single-particle energy of this orbital by 2 MeV. 
 With this modification, however, the calculated $3^-$ states become higher than those of the $fpg_{9/2}$ calculations. On the other way, it might be also possible to adjust the cross-shell pairing two-body matrix elements (TBMEs) $<(fp)^2|V|(g_{9/2}d_{5/2})^2>$ ($J=0$) to reduce the binding energy of $0_1^+$. The $3^-$ problem remains open question. 
Results corresponding to the modified single-particle energy shows that $g_{9/2}$ orbital is very crucial for heavier Ca isotopes.
The significant increase of occupancy for the $g_{9/2}$ orbital is obtained
above $N=34$ once we move towards heavier $^{54-58}$Ca isotopes. The
calculations predict that the $d_{5/2}$ orbital also plays an important role
for heavier $^{54-58}$Ca, while it is marginal for $^{47-52}$Ca.
 Our calculations support $N=32$ and $N=34$ subshell closures in the Ca isotopes for $^{52}$Ca and $^{54}$Ca, and also confirmed by the wavefunctions of $^{53}$Ca for ground state, first and second excited states. The results for the IM-SRG  interaction targeted for a particular nucleus 
with chiral $NN$ and $3N$ forces are also reported.

  \section*{ACKNOWLEDGEMENTS}

B. Bhoy acknowledges financial support from MHRD, Government
of India. PCS acknowledges a research grant from SERB (India), CRG/2019/000556.
We have performed theoretical calculations at Prayag 5 node computational facility at IIT-Roorkee. PCS would like to thank  T. Togashi and R. Stroberg for useful help.



\begin{thebibliography}{10}
\providecommand{\url}[1]{\texttt{#1}}
\providecommand{\urlprefix}{URL}
\providecommand{\eprint}[2][]{\url{#2}}

\bibitem{nature_ruiz}  R.F. Garcia Ruiz et al., Unexpectedly large charge radii of neutron-rich calcium isotopes,
{\color{blue}  Nat. Phys.  {\bf 12}, 594 (2016).}

\bibitem{holtdrip} J. D. Holt, S. R. Stroberg, A. Schwenk, and J. Simonis, Ab initio limits of atomic nuclei, {\color{blue} arXiv:1905.10475.}


\bibitem{nadya} N. A. Smirnova, B. Bally, K. Heyde, F. Nowacki, K. Sieja, Shell evolution and nuclear forces, {\color{blue}  Phys.Lett. B {\bf 686}, 109 (2010).}

\bibitem{022501}  O.B. Tarasov et al., 
Discovery of $^{60}\mathrm{Ca}$ and Implications For the Stability of $^{70}\mathrm{Ca}$, {\color{blue}  Phys. Rev. Lett. {\bf 121}, 022501 (2018).}

\bibitem{032502}  G. Hagen, M. Hjorth-Jensen, G.R. Jensen and R. Machleidt, and T. Papenbrock, 
Evolution of Shell Structure in Neutron-Rich Calcium Isotopes, {\color{blue} Phys. Rev. Lett. {\bf 109}, 032502 (2012).}

\bibitem{forssen} C. Forssen, G. Hagen, M. Hjorth-Jensen, W. Nazarewicz, and J. Rotureau, 
Living on the edge of stability, the limits of the nuclear landscape, {\color{blue} Phys. Scr. {\bf T152}, 014022 (2013).}

\bibitem{132501} G. Hagen, P. Hagen, H.-W. Hammer, and L.Platter, Efimov Physics Around the Neutron-Rich $^{60}$Ca Isotope, {\color{blue} Phys. Rev. Lett. {\bf 111}, 132501 (2013).}

\bibitem{022506}S. Michimasa et al., Magic Nature of Neutrons in $^{54}\mathrm{Ca}$: First Mass Measurements of $^{55-57}\mathrm{Ca}$, 
{\color{blue} Phys. Rev. Lett. {\bf 121}, 022506 (2018).}


\bibitem{072502} H.N. Liu et al., How robust is the $N=34$ subshell closure ? first spectroscopy of $^{52}$Ar, {\color{blue} Phys. Rev. Lett. {\bf 122}, 072502 (2019).}

\bibitem{Holt2} J. D. Holt, J. Menendez, J. Simonis, and A. Schwenk,  Three-nucleon forces and spectroscopy of neutron-rich calcium isotopes, {\color{blue}  Phys. Rev. C {\bf 90}, 024312 (2014).}

\bibitem{Lenzi} S. M. Lenzi et al.,  Island of inversion around $^{64}$Cr, {\color{blue} Phys. Rev. C {\bf 82}, 054301 (2010).}

\bibitem{keneko} K Kaneko, Y. Sun, M. Hasegawa and T. Mizusaki, Shell model study of single-particle and collective structure in neutron-rich Cr isotopes,
{\color{blue} Phys. Rev. C {\bf 78}, 064312 (2008).}

\bibitem{Caurier} E. Caurier, F. Nowacki, and A. Poves, Large-scale shell model calculations for exotic nuclei,
{\color{blue} Eur. Phys. J. A {\bf 15}, 145-150 (2002).}

\bibitem{wimmer} K. Wimmer  et al., First spectroscopy of $^{61}$Ti and the transition to the island of inversion at $N=40$, {\color{blue} Phys. Lett. B  {\bf 792}, 16 (2019).}

\bibitem{Holt1} J. D. Holt, T. Otsuka, A. Schwenk and T. Suzuki,  Three-body forces and shell structure in calcium
isotopes, {\color{blue} J. Phys. G: Nucl. Part. Phys. {\bf 39}, 085111 (2012).}



\bibitem{Stroberg} S. R. Stroberg, A. Calci, H. Hergert, J. D. Holt, S. K. Bogner, R. Roth and A. Schwenk,  Nucleus-Dependent Valence-Space Approach to Nuclear Structure,  {\color{blue}  Phys. Rev. Lett. 
{\bf 118}, 032502 (2017).}


\bibitem{Kshell} Noritaka Shimizu, Nuclear shell-model code for massive parallel computation, KSHELL (private communication).







\bibitem{Tomoaki} T. Togashi et al., Large-scale shell-model calculations for unnatural-parity high-spin states in neutron-rich Cr 
and Fe isotopes, {\color{blue} Phys. Rev. C {\bf 91}, 024320 (2015).}

\bibitem{steppenbeck} D. Steppenbeck et al., Evidence for a new nuclear magic number from the level structure of $^{54}$Ca,
{\color{blue}  Nature {\bf 502}, 207 (2013).}

\bibitem{Honma2} M. Honma et al., Shell-model description of neutron-rich Ca isotopes, {\color{blue} RIKEN Accel. Prog. Rep. {\bf 41}, 32 (2008).}

\bibitem{Honma1} M. Honma, T. Otsuka, B. A. Brown, and T. Mizusaki, Shell-model description of neutron-rich pf-shell nuclei with a
new effective interaction GXPF1, {\color{blue} Eur. Phys. J. A {\bf 25}, 499 (2005).}

\bibitem{vmu}
T. Otsuka, T. Suzuki, M. Honma, Y. Utsuno, N. Tsunoda, Novel Features of Nuclear Forces and Shell Evolution in Exotic Nuclei, 
K. Tsukiyama, and M. Hjorth-Jensen, {\color{blue}  Phys. Rev. Lett. {\bf 104},
012501 (2010)}.

\bibitem{GL} D. H. Gloeckner and R. D. Lawson, Spurious center-of-mass motion, {\color{blue}  Phys. Lett. 53B, 313 (1974)}.

\bibitem{pcs_mn} P.C.Srivastava and I.Mehrotra, Large-scale shell model calculations for odd-odd $^{58-62}$Mn isotopes,
 {\color{blue} Eur. Phys. J. A {\bf 45} , 185 (2010).}


\bibitem{machleidt11_n3lo} R. Machleidt and D.R. Entem, Chiral effective field theory and nuclear forces,  {\color{blue}  Phys. Rep.
{\bf 503}, 1 (2011).}

\bibitem{machleidt12_n2lo} D.R. Entem and R. Machleidt, Accurate charge-dependent nucleon-nucleon potential at fourth order of chiral
perturbation theory, {\color{blue} Phys. Rev. C {\bf 68}, 041001(R) (2003).}

\bibitem{navratil} P. Navratil, Local three-nucleon interaction from chiral effective field theory, {\color{blue}  
Few-Body Syst.  {\bf 41}, 117 (2007).}



\bibitem{Huck} A. Huck, G. Klotz, A. Knipper, C. Miehe, C. Richard-Serre, G. Walter, A. Poves, H. L. Ravn, and G. Marguier, 
Beta decay of the new isotopes $^{52}$K, $^{52}$Ca, and $^{52}$Sc; a test of the shell model far from stability, {\color{blue}
Phys. Rev. C {\bf 31}, 2226 (1985).}

\bibitem{NNDC} {\color{blue} http://www.nndc.bnl.gov/ensdf/}.


\bibitem{montanari} D. Montanari, S. Leoni, D. Mengoni, J. J. Valiente-Dobon,
G. Benzoni, N. Blasi, G. Bocchi, P. F. Bortignon, S. Bottoni, and
A. Bracco, $\gamma$ spectroscopy of calcium nuclei around doubly magic $^{48}$Ca using heavy-ion transfer reactions, {\color{blue} 
Phys. Rev. C {\bf85}, 044301 (2012).}

\bibitem{Raman} S. Raman, C. W. G. Nestor Jr., and P. Tikkanen, Tables of E2 transition probabilities from the first 2$^{+}$ states in
even-even nuclei, {\color{blue} At. Data Nucl. Data Tables {\bf78}, 1 (2001).}





\bibitem{Crawford} H. L. Crawford, A. O. Macchiavelli, P. Fallon, M. Albers, Unexpected distribution of $\nu$1f$_{7/2}$ strength in $^{49}$Ca, 
{\color{blue} Phys. Rev. C {\bf95}, 064317 (2017).}




\bibitem{ruiz} R.F. Garcia Ruiz et al.,  Ground-state electromagnetic moments of calcium isotopes, {\color{blue} Phys. Rev. C {\bf 91}, 041304(R) (2015).}


\bibitem{PRL_nav}  S. Chen et al.,  Quasifree neutron knockout from $^{54}$Ca corroborates arising $N = 34$ neutron
magic number, {\color{blue} Phys. Rev. Lett. {\bf 123}, 142501  (2019).}

\bibitem{gazit} D. Gazit, S. Quaglioni, and P.  Navratil, Erratum: Three-Nucleon Low-Energy Constants from the Consistency
of Interactions and Currents in Chiral Effective Field Theory  {\color{blue} Phys. Rev. C {\bf 122}, 029901(E) (2019).}



\end{thebibliography}
\end{document}